\documentstyle[12pt]{article}

\newlength{\extraspace}
\newlength{\extraspaces}
\newcommand{\lae}{\begin{array}{c}\,\sim\vspace{-21pt}\\< \end{array}}

\newcommand{\be}{\begin{equation}}
\newcommand{\ee}{\end{equation}}
\newcommand{\bear}{\begin{eqnarray}}
\newcommand{\eear}{\end{eqnarray}}
\newcommand{\ba}{\begin{array}}
\newcommand{\ea}{\end{array}}

\newcommand{\pq}{\mbox{$f_{\rm PQ}$}}  
\begin{document}
\pagestyle{empty}
\begin{titlepage}

\makebox[10.4cm][l]{BUHEP-96-30} hep-ph/9609221

\makebox[10.4cm][l]{Submitted to Phys.\ Rev.\ {\bf D}} September 1, 1996

\makebox[10.4cm][l]{} Revised October 21, 1996

\vspace{24pt}
\begin{center}
{\LARGE {\bf The strong CP problem versus
\\ [3mm] Planck scale physics    }}\\
\vspace{40pt}
{\large {\rm Bogdan A. Dobrescu}}\footnote{e-mail address:
dobrescu@budoe.bu.edu}
\vspace*{0.5cm}

{\it Department of Physics, Boston University \\
590 Commonwealth Avenue, Boston, MA 02215, USA}

\vskip 2.8cm
\end{center}
\baselineskip=18pt

\begin{abstract}

We discuss conditions that should be satisfied by
axion models for solving the strong CP problem.
It has been observed that Planck scale effects
may render the axion models ineffective if there are
gauge invariant operators of dimension less than 10 which break
explicitly the Peccei-Quinn (PQ) symmetry.
We argue that only those operators formed
of fields which have vacuum expectation values are dangerous.
Supersymmetric axion models fail to prevent even this restricted 
class of operators. Furthermore, the models that relate the PQ
scale and the supersymmetry breaking scale are particularly sensitive
to gauge invariant PQ-breaking operators.
By contrast, in non-supersymmetric composite axion models
the PQ scale arises naturally, and the dangerous operators can
be avoided. However, the composite axion models contain 
heavy stable particles which are cosmologically ruled out.
Another problem is a Landau pole for the QCD coupling constant. 
Both these problems may be solved if the unification of color 
with the gauge interactions which bind the axion could be achieved.

\end{abstract}

\vfill
\end{titlepage}

\baselineskip=18pt
\pagestyle{plain}
\setcounter{page}{1}

%%%%%%%%%%%%%%%%%%%%%%%%%%%%%%%%%%%%%%%%%%%%%%%%%%%%%%%%%%%%%%%%
\section{Introduction}
\label{sec:intro}
\setcounter{equation}{0}

The fine tuning required to accommodate the 
observed CP invariance of the strong interactions, known as the
strong CP problem \cite{cheng},
suggests that the strong CP parameter $\bar{\theta}$ is
a dynamical field.
If some colored fields are charged
under a spontaneously broken Peccei-Quinn (PQ) symmetry \cite{pq},
then $\bar{\theta}$ is replaced by a shifted axion field.
The PQ symmetry is explicitly broken by QCD instantons, so that
a potential for the axion is generated with a minimum at $\bar{\theta}
= 0$. In the low energy theory, besides solving the strong CP problem,
this mechanism predicts nonderivative couplings
of the axion to the gauge bosons and model dependent  derivative
couplings of the axion to hadrons and leptons \cite{eff,hadronic}.

There are two important issues that have to be addressed 
by axion models. First, Planck scale effects may break explicitly the
PQ symmetry, shifting $\bar{\theta}$ from the origin
\cite{planck,ten,effect}.
Since only the gauge symmetries are expected to be preserved by Planck
scale physics \cite{grav}, 
the PQ symmetry should be a consequence of a gauge symmetry.

Second, an axion model should produce naturally the PQ symmetry
breaking scale, \pq.
Astrophysics and cosmology \cite{astro} constrain the axion mass
to lie between $10^{-5}$ and $10^{-3}$ eV \cite{data},
which translates in a range $10^{10}-10^{12}$ GeV for \pq.
The small ratio between the PQ scale and the Planck scale, $M_{\rm P}
\sim 10^{19}$ GeV,
can be naturally explained if the PQ symmetry is broken dynamically
in a theory with only fermions and gauge fields \cite{comp, model}. 
Alternatively, if the PQ symmetry is broken by the vacuum expectation
value (vev) of a fundamental scalar, then supersymmetry (susy)
is required to protect \pq\ against quadratic divergences.

In this paper we study phenomenological constraints on axion models
and point out potential problems of the models
constructed so far.
In section \ref{sec:prot} 
we discuss under what conditions a gauge symmetry can protect 
$\bar{\theta}$ from Planck scale effects. We also list 
theoretical and phenomenological requirements that should be imposed
on axion models.
These conditions are illustrated in the case of
non-supersymmetric composite axion models 
in section \ref{sec:comp}. In section \ref{sec:susy} 
it is shown that previous attempts of preventing harmful
PQ breaking operators in supersymmetric theories have failed. 
A discussion of the PQ scale in supersymmetric models is also included.
Conclusions and a summary of results are presented 
in section \ref{sec:conc}. 

\vfil
\newpage

%%%%%%%%%%%%%%%%%%%%%%%%%%%%%%%%%%%%%%%%%%%%%%%%%%%%%%%%%%%%%%%%
\section{Constraints on axion models}
\label{sec:prot}
\setcounter{equation}{0}

\subsection{Protecting the axion against Planck scale effects}

Gravitational interactions are expected to break any continuous
or discrete global symmetries \cite{grav},
so that gauge invariant nonrenormalizable operators suppressed by
powers of $M_{\rm P}$ are likely to have coefficients of order 
one.
In refs.~\cite{ten,effect} it is argued that, under these
circumstances, a solution to the strong CP problem requires any gauge 
invariant operator of dimension less than 10 to preserve the PQ
symmetry.
The reason is that the PQ-breaking operators change the potential
for the axion such that the minimum moves 
away from $\bar{\theta} = 0$.
However, this condition can be relaxed. If a PQ-breaking operator 
involves fields which do not have vevs, then its effect is an
interaction of the axion with these fields.
The exchange of these fields will lead to a potential for the
axion which is suppressed by at least as many powers of $M_{\rm P}$
as the lowest dimensional PQ-breaking operator formed by fields
which have vevs. Therefore, a natural solution to the strong CP
problem requires that {\it gauge symmetries forbid any PQ-breaking
  operator of dimension less than 10 involving only fields which
  acquire vevs of order \pq.}

This relaxed form is still strong enough to raise the question
of whether we should worry that much about Planck scale effects 
which are mostly unknown.
Furthermore, in ref.~\cite{wormhole} it is argued that although
the idea of wormhole-induced global symmetry breaking is robust,
some modifications of the theory of gravity at a scale of $10^{-1}
M_{\rm P}$ or topological effects in string theory
could lead to exponentially suppressed coefficients
of the dangerous operators. There are also arguments that strongly 
coupled heterotic string theory may contain PQ symmetries which are 
adequately preserved \cite{schstring}.

Nevertheless, since the theory of quantum gravity still eludes us,
assigning exponentially small coefficients to all the gauge invariant
PQ-breaking  operators in the low energy theory can be seen as a
worse fine-tuning than setting $\bar{\theta} < 10^{-9}$.
To show this consider a scalar $\Phi$, charged under a global 
U(1)$_{\rm PQ}$ which has a QCD anomaly, with a vev equal to \pq,
and a dimension-$k$ gauge invariant operator 
\be
\frac{c}{k!}
\frac{1}{M_{\rm P}^{k - 4}} \Phi^k ~,
\ee
where $c$ is a dimensionless coefficient.
Solving the strong CP problem requires
\be
\frac{c}{k!}
\frac{f^k_{\rm PQ}}{M_{\rm P}^{k - 4}} < \bar{\theta} \,
M^2_{\pi} f^2_{\pi} ~.
\label{first}
\ee
Here $M_{\pi}$ is the pion mass, and $f_{\pi} \approx 93$ MeV is 
the pion decay constant.
Therefore, the condition on $c$ is
\be
|c| \lae \bar{\theta} \, k!\, 10^{8 (k - 10)}
\left(\frac{10^{11} \, {\rm GeV}}{\pq}\right)^{\!\! k}~,
\ee
which means that 
$c$ is less finely tuned than $\bar{\theta}$ only if $k \ge 9$
($k \ge 11$) 
for $\pq = 10^{10}$ GeV ($\pq = 10^{12}$ GeV).

%%%%%%%%%%%%%%%%%%%%%%%%%%%%%%%%%%%%%%%%%%%%%%%%%%%%%%%%%%%%%%%%%
\subsection{General conditions}

Even in its relaxed form, the condition of avoiding Planck scale
effects is hard to satisfy simultaneously with the other
requirements of particle physics, cosmology and astrophysics.
In the remainder of this section we list some of the important
issues in axion model building.

\noindent
i) Gauge anomaly cancellation.

\noindent
ii) The colored fields carrying PQ charges should not acquire
vevs which break SU(3)$_C$ color.

\noindent
iii) The stability of \pq\ requires either susy or
the absence of fundamental scalars at this scale.
Furthermore, any mass parameter except $M_{\rm P}$ should arise
from the dynamics. Otherwise, fine-tuning 
the ratio $\pq/M_{\rm P}$ is as troublesome as imposing 
$\bar{\theta} < 10^{-9}$, and the motivation for axion models 
is lost. Note that the usual DFSZ \cite{dfsz} and KSVZ \cite{ksvz}
models do not satisfy this condition.

\noindent
iv) The strong coupling constant should remain small above \pq,
until $M_{\rm P}$ or some grand unification scale. 
The one-loop renormalization group evolution for the strong
coupling constant, starting with 5 flavors from
$\alpha_s(M_Z) = 0.115$, then at 175 GeV including the top quark,
gives
\be
\frac{1}{\alpha_s(\pq)} \approx
32.0 + \frac{7}{2 \pi} \log\left(\frac{\pq}
{10^{11} \, {\rm GeV}}\right) ~.
\label{rge}
\ee
Running then $\alpha_s$ from \pq\ to $M_{\rm P}$ gives 
$\alpha_s(M_{\rm P}) < 1$ if the coefficient of the $\beta$ function
is $b_0 \lae 10.6$. This corresponds to a maximum of 26 new flavors.
In supersymmetric theories the above computation gives
$\alpha_s(\pq) \approx 1/19$, and $\alpha_s(M_{\rm P}) 
< 1$ if $b_0 \lae
6$, i.e. there can be at most 10 new flavors with masses of order
\pq. If there are additional flavors below \pq, the total number of
flavors allowed is reduced. In the case of composite axion models
there are non-perturbative effects, due to the fields carrying
the confining gauge interactions and the usual color, 
which change the running of $\alpha_s$ at scales close to \pq\ and
can be only roughly estimated.

\noindent
v) Composite stable particles with masses $M_{\rm comp}$ 
larger than about $10^5$ GeV
lead too early to a matter dominated universe  \cite{bbf}.
It is then necessary that all stable 
particles with masses of order \pq\ to be short
lived. Their energy density remains smaller than the critical one
provided their lifetime is smaller than of order  $10^{-8}$ 
seconds \cite{decay2}. 
However, if there is inflation any unwanted relic is
wiped out, and if the reheating temperature is lower than \pq,
then the heavy stable particles are not produced again and
the above condition is no longer necessary.

\noindent
vi) Domain walls may arise in many axion models \cite{wall},
and they should 
disappear before dominating the matter density of the universe.
Inflation takes care of this requirement too, but there are 
also other mechanisms for allowing the domain walls to evaporate
\cite{cheng,effect,anomaly}. 

\noindent
vii) Any new colored particle should be heavier than about the
electroweak scale \cite{data}. Etc.

%%%%%%%%%%%%%%%%%%%%%%%%%%%%%%%%%%%%%%%%%%%%%%%%%%%%%%%%%%%%%%%%%
\section{Composite axion}
\label{sec:comp}
\setcounter{equation}{0}

The PQ scale is about 9 orders of magnitude smaller than the Planck
scale,
which is unnatural unless the spontaneous breaking of the PQ symmetry
is a consequence of non-perturbative effects of some non-Abelian gauge
symmetry. In this section we concentrate on non-supersymmetric
theories, and therefore we do not allow light (compared to $M_{\rm
  P}$) fundamental scalars. 
We have to consider then theories with fermions transforming
non-trivially under a gauge group. From QCD it is known that 
the strong dynamics break the chiral symmetry of the quarks.
Thus, if the PQ symmetry is a subgroup of a chiral symmetry
in a QCD-like theory, then \pq\ will be of the order of the
scale where the gauge interactions become strong.
As a result the axion will be a composite state, formed of the
fermions charged under the confining gauge interactions.

%%%%%%%%%%%%%%%%%%%%%%%%%%%%%%%%%%%%%%%%%%%%%%%%%%%%%%
\subsection{Kim's model}

The idea of a composite axion is explicitly realized in
the model presented in ref.~\cite{comp},
which contains fermions carrying color and the charges of an SU(N)
gauge interaction, 
called axicolor. The left-handed fermions are in the
following representations of the SU(N)$\times$SU(3)$_C$ gauge group:
\be
\psi: (N,3) \, , \; \phi: (N,1) \, , \; \chi:
(\overline{N},\bar{3}) \, , \; \omega: (\overline{N},1) ~.
\ee
SU(N) becomes strong at a scale $\Lambda_{\rm a}$ of order \pq\
and the fermions condense.
This is a QCD-like theory with $N$ axicolors and 4 flavors,
and from QCD we know that the condensates will preserve SU(N).
In the limit where the SU(3)$_C$ coupling constant, $\alpha_s$,
 is zero, the channels of condensation which preserve color are 
equally attractive as the ones which break color. 
Thus, although $\alpha_s$ is small at the scale $\Lambda_{\rm a}$, 
its non-zero value will force the condensates to preserve color,
which implies that only the $\langle \psi\chi \rangle$ and
$\langle \phi\omega \rangle$ condensates will form.

In the limit $\alpha_s \rightarrow 0$, Kim's model has an 
SU(4)$_L\times$SU(4)$_{\overline{R}}\times$U(1)$_{\rm V}$
global symmetry which is spontaneously broken down to
SU(4)$_{L-\overline{R}}\times$U(1)$_{\rm V}$ by the condensates. 
The resulting 15 Goldstone bosons transform as $1 + 3 + \bar{3} + 8$
under SU(3)$_C$.
The color singlet is the composite axion, with a ($\psi\chi - 3
\phi\omega$) content, and U(1)$_{\rm PQ}$ corresponds to the 
$[(Q_{\rm PQ})_L \times 1_{\overline{R}} + 1_L \times (Q_{\rm
  PQ})_{\overline{R}}]/\sqrt{2}$
broken generator of 
SU(4)$_L\times$SU(4)$_{\overline{R}}$, where 
\be
Q_{\rm PQ} = \frac{1}{2\sqrt{6}} {\rm diag} (1,1,1,-3) ~.
\ee

When $\alpha_s$ is turned on, the SU(4)$_{L-\overline{R}}$ 
global symmetry is explicitly broken down to the gauged SU(3)$_C$
and the global U(1)$_{{\rm axi-} B-L}$ generated by 
$(Q_{\rm PQ})_{L-\overline{R}}$. 
The axion gets a tiny mass from QCD
instantons, while the other (pseudo) Goldstone bosons get masses 
from gluon exchange. 

Although the normalization of the PQ symmetry breaking scale, \pq, is
ambiguous, 
the axion mass, $m_{\rm a}$, is non-ambiguously related to the
axicolor scale, $\Lambda_{\rm a}$,
because U(1)$_{\rm PQ}$ is a subgroup of the chiral symmetry 
of axicolor. To find this relation note first that 
the axion mass is determined by the ``axi-pion'' decay constant,
$f_a$ (the analog of $f_{\pi}$ from QCD), by \cite{hadronic}
\be
m_{\rm a} = \frac{4 A_{\rm PQ}^C}{f_{\rm a}} M_{\pi} f_{\pi} 
\frac{Z^{1/2}}{1 + Z} ~,
\ee
where $Z \approx 0.5$ is the up to down quark mass ratio,
and $A_{\rm PQ}^C$ is the color anomaly of U(1)$_{\rm PQ}$:
\be
\delta_{ab} A_{\rm PQ}^C = N \, {\rm Tr} (T_a T_b Q_{\rm PQ}) ~.
\ee
The normalization of the SU(3)$_C$ generators [embedded in 
SU(4)$_{L-\overline{R}}$] is ${\rm Tr} (T_a T_b) 
= \delta_{ab}/2$, and we find
\be
f_{\rm a} = 2.4 \times 10^9 \,{\rm GeV}
\left(\frac{10^{-3} \,{\rm eV}}{m_{\rm a}}\right) N ~. 
\ee
In the large-$N$ limit the relation between $f_{\rm a}$ and 
$\Lambda_{\rm a}$ is
\be
\frac{\Lambda_{\rm a}}{\Lambda_{\rm QCD}} = \frac{f_{\rm a}}{f_{\pi}} 
\sqrt{\frac{3}{N}} ~,
\label{largeN}
\ee
where $\Lambda_{\rm QCD} \sim 200$ MeV.

This model suffers from the energy density problem of 
stable composite particles \cite{decay1} [see point v) in section 2].
The reason is that the global U(1)$_{\rm V}$
(the analog of the baryon number symmetry in QCD) 
is an exact symmetry such that the lightest axibaryon is stable.
Its mass is larger than $f_{\rm a}$ and 
can be evaluated as in ref.~\cite{tcbar} by scaling from QCD:
\be
M_{\rm aB} = 
m_p \left(\frac{f_{\rm a}}{f_{\pi}}\right) \sqrt{\frac{N}{3}} ~,
\label{abar}
\ee
where $m_p$ is the proton mass.
If axicolor can be unified with a standard model gauge group, then
the heavy gauge bosons would mediate the decay of the axibaryons
into standard model fermions and the model would be cosmologically safe
\cite{decay2}. However, it will be highly non-trivial to
achieve such a unification. The only attempt so far of avoiding the
axibaryon cosmological problem 
involves scalars  \cite{decay1}, 
so it is unsatisfactory unless one shows that
these scalars can be composite states. 

We point out that the axibaryons are not the only heavy stable
particles: the color triplet 
pseudo Goldstone bosons (PGB's) have also a too large energy density. 
Their masses can be estimated by scaling
the contribution from electromagnetic interactions to the pion mass,
which is related to the difference between the squared masses of
$\pi^{\pm}$ and $\pi^0$. Since $\alpha_s(\Lambda_{\rm a})$ 
is small, the bulk of the colored PGB's masses comes from 
one gluon exchange \cite{tc}:
\be
M^2_{(R)} \approx C^2(R)
\frac{\alpha_s(\Lambda_{\rm a})}{\alpha(\Lambda_{\rm QCD})}
 \frac{\Lambda_{\rm a}^2}{ \Lambda_{\rm QCD}^2}
\left(M^2_{\pi^{\pm}} - M^2_{\pi^0}\right) ~.
\label{pgb}
\ee
Here\footnote{Eq.~(\ref{pgb}) improves the estimate given 
in \cite{decay1,tcs} by eliminating the dependence on $N$
shown in eq.~(\ref{largeN}).}
$R$ is the SU(3)$_C$ representation, 
$C^2(R)$ is the quadratic Casimir, equal to 3 for the color
octet and 4/3 for the triplet, and
$\alpha$ is the electromagnetic coupling constant.
Therefore, the color triplet PGB's , which are
$\psi\omega$ and $\phi\chi$ bound states, have a mass
\be
M_{(3, \bar{3})} \approx 0.9 f_{\rm a} \sqrt{\frac{3}{N}} 
\label{triplet}
\ee
and, except for the axion, are the lightest ``axihadrons''. 
These are 
absolutely stable due to the exact global U(1)$_{{\rm axi-}(B-L)}$
symmetry. 
One may choose though not to worry about stable axihadrons
by assuming a period of inflation with reheating temperature
below the PGB mass.

The model discussed so far does not attempt to avoid the 
Planck scale induced operators which violate the PQ symmetry.
In fact, Kim's model is vector-like: the $\psi$ and $\chi$, 
as well as the  $\phi$ and $\omega$, will pair to form Dirac 
fermions. Their mass is likely to be of order $M_{\rm P}$
and then fermion condensation does not take place and 
the model becomes useless.
Even if Planck scale masses for the fermions are not generated,
there are dimension 6 operators which violate U(1)$_{\rm PQ}$: 
\be
\frac{c_1}{M_{\rm P}^2}(\psi\chi)^2 \;\; , \;\;  
\frac{c_2}{M_{\rm P}^2}(\phi\omega)^2 ~,
\ee
where $c_j$, $j = 1, 2$, are dimensionless coefficients.
These operators will shift the vev of the axion such that
$\bar{\theta}$ will remain within the experimental bound 
only if 
\be
\frac{9|c_1| + |c_2|}{M_{\rm P}^2}
\left(4 \pi f_{\rm a}^3 \sqrt{\frac{3}{N}}\, \right)^{\! 2}
< 10^{-9} M^2_{\pi} f^2_{\pi} ~,
\ee
implying $|c_j| < {\cal O}(10^{-47})$. 
It is hard to accept this tiny number given that
the motivation for studying axion models is 
to explain the small value $\bar{\theta} < 10^{-9}$. 

%%%%%%%%%%%%%%%%%%%%%%%%%%%%%%%%%%%%%%%%%%%%%%%%%%%%%%%%%%%%%%%%%%
\subsection{Randall's model}

There is only one axion model in the literature which does not involve
scalars and avoids large Planck scale effects \cite{model}.
To achieve this, Randall's model includes 
another gauge interaction, which is weak, in addition to
the confining axicolor. 
The left-handed fermions transform under
the SU(N)$\times$SU(m)$\times$SU(3)$_C$ gauge group as: 
\be
\psi: (N,m,3) \, , \; \phi_i: (N,\overline{m},1) \, , \; \chi_j:
(\overline{N},1,\bar{3}) \, , \; \omega_k: (\overline{N},1,1) ~,
\ee
where $i=1,2,3$, $j=1,...,m$, and $k=1,...,3 m\,$ are flavor indices.
Axicolor SU(N) becomes strong at the $\Lambda_{\rm a}$ 
scale and the fermions condense.
If the SU(m) gauge coupling, $g_m$, is turned off, the vacuum will
align as in Kim's model and will preserve color.
When $g_m$ is non-zero, the SU(m) gauge interaction will 
tend to change the vacuum alignment and break the SU(N) gauge
symmetry.
However, since $g_m$ is small, this will not happen,
as we know from QCD where the weak interactions of the quarks do not
affect the quark condensates. Therefore, the 
\be
\frac{1}{3} \langle \psi \chi_j \rangle = 
\langle \phi_i \omega_k \rangle \approx 
4 \pi f_{\rm a}^3 \sqrt{\frac{3}{N}} ~.
\ee
condensates are produced, breaking the SU(m) gauge group and
preserving color. 
A global U(1)$_{\rm PQ}$, under which $\psi$ and $\chi_j$ have
charge +1 while $\phi_i$ and $\omega_k$ have charge $-1$, is
spontaneously broken by the condensates, so that an axion arises.

The lowest dimensional gauge invariant and PQ-breaking 
operators involving only fields that acquire vevs are
\be 
c^{ijk}_{m^{\prime}}
\frac{1}{M_{\rm P}^{3 m - 4}} 
\left(\psi \chi_j\right)^{m - m^{\prime}}
(\overline{\phi_i} \overline{\omega_k})^{m^{\prime}}~,
\ee
with $m^{\prime} = 1,...,m$ ($m^{\prime} \neq m/2$). The
$c^{ijk}_{m^{\prime}}$ coefficients are assumed to be of order one.
The solution to the strong CP problem requires
\be
\frac{C(m)}{M_{\rm P}^{3 m - 4}}
\left(4 \pi f_{\rm a}^3 \sqrt{\frac{3}{N}}\, \right)^{\!\! m}
< 10^{-9} M^2_{\pi} f^2_{\pi} ~,
\label{ineq}
\ee
where
\be
C(m) \equiv \sum\limits_{ijkm^{\prime}} 3^{m - m^{\prime}}
\left| c^{ijk}_{m^{\prime}} \right| ~.
\ee
A necessary condition that follows from inequality (\ref{ineq}) is
$3 m \geq 10$.
Note that the window $\pq \sim 10^7$ GeV discussed in \cite{model}
has been closed \cite{ressell}.
This constraint on $m$,
combined with the condition of asymptotic freedom for SU(N)
gives a lower bound for $N$,
\be
\frac{11}{12} N > m \ge 4 ~.
\label{integ}
\ee
We will see shortly that $m = 4$, $N = 5$ are the only values that may
not lead to a Landau pole for QCD much below $M_{\rm P}$.
For these values of $m$, inequality (\ref{ineq}) yields an upper limit
for the axi-pion decay constant:
\be
f_{\rm a} < \frac{1.9 \times 10^{11} \, {\rm GeV}}{C(4)^{1/12}} ~.
\label{falim}
\ee
For random values of order one of the $c^{ijk}_{m^{\prime}}$
coefficients, we expect $C(4)^{1/12}$ to be between 1.5 and 2.5.
For example, if $c^{ijk}_{m^{\prime}}$ = 1,
then $C(m) = 9 m^2 (3^{m + 1} - 1)/2$, 
which gives $C(4)^{1/12} \approx 2.25$.
Therefore, $f_{\rm a} \lae 10^{11}$ GeV is necessary for
avoiding fine-tuning of the higher dimensional operators
in the low energy effective Lagrangian. 

We note that $m = 4$
allows dimension-9 gauge invariant operators which
break U(1)$_{\rm PQ}$:
\be
(\overline{\phi_i}\psi)^2 (\overline{\chi_j} \omega_k) \, , \; 
(\psi\omega_k)^2 (\overline{\phi_i}\psi) \, , \;
(\overline{\psi}\phi_i) (\phi_l\chi_j)^2 ~.
\ee
However, these are not harmful because they are not 
formed of the fields 
which acquire vevs, i.e. $(\psi \chi_j)$ and $(\phi_i \omega_k)$.
They will just induce suppressed interactions of the axion with the
fermions. Hence, this model is an example where the redundant 
condition of avoiding {\it all} the operators of dimension less than
10 is not satisfied.

Randall's model has a non-anomalous 
SU(3m)$\times$SU(m)$\times$SU(3)$\times$U(1)$_{{\rm axi-}(B-L)}$ 
global symmetry under which the fermions transform as:
\be
\psi: (1,1,1)_{+1} \, , \; \phi: (1,1,3)_{-1} \, , \; \chi:
(1,m,1)_{-1}  \, , \; \omega: (3 m,1,1)_{+1} ~.
\ee
 
This  global symmetry, combined
with the SU(m) gauge symmetry, is
 spontaneously broken down to an [SU(m)$\times$SU(3)$]_{\rm global}
\times$U(1)$_{{\rm axi-}(B-L)}$ global symmetry by the condensates.
Thus, there are $10 m^2 - 2 = 158$ Goldstone bosons: $m^2 - 1$ 
of them are eaten by the SU(m) gauge bosons
which acquire a mass $g_m \pq/2$, while the other $9 m^2 - 1$ 
get very small masses from higher dimensional
operators. These are color singlets,
very weakly coupled to the standard model
particles, and, as pointed out in \cite{model},
their energy density might not pose cosmological problems.

The Goldstone bosons have $\psi\chi$ and $\phi\omega$ content 
and transform in the $(m,1)$ and $(m,3)$ representations
of the unbroken [SU(m)$\times$SU(3)$]_{\rm global}$,
respectively. Therefore, 
these symmetries do not prevent heavy resonances from decaying
into Goldstone bosons, and are cosmologically safe. However,
as in Kim's model, the lightest particles carrying 
axi-$(B-L)$ number are the color triplet PGB's, 
which have $\psi\omega_k$ and 
$\phi_i\chi_j$ content and are heavy due to gluon exchange.
Hence, there are $18 m^2$ stable ``aximesons''  
with masses given by eq.~(\ref{triplet}), which 
pose cosmological problems.

Besides heavy stable particles, there are meta-stable 
states, with very long lifetimes, incompatible with the thermal
evolution of the early universe. To show this we observe
that there is an axibaryon
number symmetry, U(1)$_{\rm V}$, broken only by the SU(m) anomaly.
The $\psi$ and $\chi$ fermions have U(1)$_{\rm V}$ charge +1
while $\phi$ and $\omega$ have charge $-1$.  
The lightest axibaryons are the color singlet 
$\, \psi^{3 p_1}\phi^{3 p_2}\chi^{p_3}\omega^{p_4}\, $ 
states \cite{tcbar}, 
with $p_l \ge 0$ ($l = 1,...,4$) integers satisfying
$\sum p_l = N$.
These can decay at low temperature 
only via SU(m) instantons, with a rate proportional to $\exp(- 16
\pi^2/g_m^2)$, which is extremely small given that the SU(m) gauge
coupling is small.
At a temperature of order \pq\ transitions between vacua with
different axibaryon number via sphalerons  will affect
the axibaryon energy density. Nonetheless, this thermal effect
is exponentially suppressed as the universe cools down such that
the order of magnitude of the axibaryon energy density is unlikely 
to have time to change significantly.

As in Kim's model, unification of axicolor with other gauge groups
will allow axihadron decays if the axicolored fermions belong to the
same multiplet as some light fermions.
In this model though it seems even more difficult
to unify axicolor with other groups.
Inflation with reheating temperature below the axibaryon mass $M_{\rm
  aB}$ [see eq.~(\ref{abar})]
appears then a necessary ingredient. 

Another problem may be the existence of a large number of
colored particles. QCD not only loses asymptotic freedom
but in fact the strong coupling constant may hit the Landau pole
below $M_{\rm P}$. To study this issue we need to evaluate
$\alpha_s(M_{\rm P})^{-1}$. Below the scale set by the
mass of the PGB's, the effects of the ``axihadrons'' on the 
running of $\alpha_s$ are negligible and we can use eq.~(\ref{rge}).
Above some scale $\Lambda_{\rm pert}$ larger than 
$4 \pi f_a/\sqrt{N}$ the perturbative renormalization group evolution
can again be used, with $m N + 6$ flavors.
However, at scales between the
mass of the PGB's and $\Lambda_{\rm pert}$, besides the perturbative
contributions from the gluons and the six quark flavors, 
 there are large
non-perturbative effects of the axicolor interaction
which are hard to estimate. We can write
\be
\frac{1}{\alpha_s(M_{\rm P})} = 32.0 + \frac{7}{2 \pi}
  \log\left(\frac{M_{\rm P}} {10^{11} \, {\rm GeV}}\right) 
- \frac{m N}{3 \pi} 
\log\left(\frac{M_{\rm P}}{\Lambda_{\rm pert}}\right)
 - \delta_{\rm PGB} - \delta_{\rm axicolor} ~.
\label{evol}
\ee 
Here $\delta_{\rm PGB}$ is the contribution from colored PGB's,
and $\delta_{\rm axicolor}$ is the non-perturbative contribution
of the axicolored fermions, which  
can be interpreted as the effect of the
axihadrons on the running of $\alpha_s$. Axicolor interactions 
have an effect on the size of
these two non-perturbative contributions, but it is unlikely that 
they change the signs of the one-loop contributions from PGB's
and axihadrons. Therefore, we expect $\delta_{\rm PGB}$ and 
$\delta_{\rm axicolor}$ to be positive.
This is confirmed by the estimate of the hadronic contributions
to the photon vacuum polarization \cite{hlm} within relativistic 
constituent quark models, and by the study of the running of 
$\alpha_s$ in technicolor theories \cite{hl}, which indicate
\be
\delta_{\rm axicolor} > \delta_{\rm PGB} > 0 ~.
\label{nonpert}
\ee
From eq.~(\ref{evol}) one then can see that $\alpha_s^{-1}(M_{\rm P})$
is negative for any  $m$ and $N$ larger than the  smallest values
allowed by eq.~(\ref{integ}): $m = 4,\, N = 5$.
With these values eq.~(\ref{evol}) becomes
\be
\frac{1}{\alpha_s(M_{\rm P})} = 13.4 + \frac{20}{3\pi}
  \log\left(\frac{\Lambda_{\rm pert}}{10^{11} \, {\rm GeV}}\right)
 - \delta_{\rm PGB} - \delta_{\rm axicolor} ~.
\label{evol1}
\ee

At low energies compared to $\Lambda_{\rm a}$, 
$\delta_{\rm PGB}$ can be evaluated
using chiral perturbation theory \cite{hlm}. Furthermore, as discussed
in ref.~\cite{hl} for the case of technicolor theories, the result
can be estimated up to a factor of 2 by computing the
one-loop PGB graphs. 
At energies larger than $\Lambda_{\rm a}$ 
chiral perturbation theory is not
useful and the contribution to $\delta_{\rm PGB}$ is unknown. Keeping
this important caveat in mind we will evaluate the one-loop
PGB contributions. The leading log term from the
$3 m^2$ color triplet PGB's 
with mass $M_{(3,\bar{3})}$ [see eq.~(\ref{triplet})]
and the $m^2$ color octet PGB's with mass $(9/4) 
M_{(3,\bar{3})}$ is given by
\be
\delta_{\rm PGB} \approx K \frac{m^2}{\pi} 
\log\left(\sqrt{\frac{2}{3}}
\frac{\Lambda_{\rm pert}}{M_{(3,\bar{3})} }\right) ~.
\label{pgbev}
\ee
$K$ is a constant between 1 and 2 which accounts for higher 
order corrections. Using eqs.~(\ref{nonpert})-(\ref{pgbev}) 
we can write
\be
\frac{1}{\alpha_s(M_{\rm P})} < 11.8 - \frac{76 }{3\pi}
\log\left(\frac{\Lambda_{\rm pert}}{f_{\rm a}}\right)
- \frac{20}{3\pi}
\log\left(\frac{10^{11} \, {\rm GeV}}{f_{\rm a}}\right) ~,
\label{con}
\ee
where we used $K = 1$.
The right-hand side of this inequality is negative
because $f_a \lae 10^{11}$ GeV [see eq.~(\ref{falim})]
and $ \Lambda_{\rm pert}/f_a
> 4\pi/\sqrt{5} $, 
which means that the strong coupling constant hits the
Landau pole below $M_{\rm P}$.

Although the estimate of the non-perturbative effects
on the RGE is debatable, this conclusion seems quite robust.  
A possible resolution would be to embed SU(3)$_C$ in a larger gauge 
group. In doing so, axicolor will lose asymptotic freedom
unless it is also embedded in the larger group. 
Such an unification of color and axicolor would solve
both problems discussed here: heavy stable particles and
the Landau pole of QCD. However, it remains to be proved
that this unification is feasible, given
the large groups already involved.

%%%%%%%%%%%%%%%%%%%%%%%%%%%%%%%%%%%%%%%%%%%%%%%%%%%%%%%%%%%%%%%
\section{Supersymmetric axion models}
\label{sec:susy}
\setcounter{equation}{0}

\subsection{Planck scale effects in supersymmetric models}

Apparently it is easier to build supersymmetric models in which 
the axion is protected against Planck scale effects because
the holomorphy of the superpotential eliminates many of the higher
dimensional operators. In practice, susy is broken so that 
the holomorphy does not ensure $\bar{\theta} < 10^{-9}$.

For example, consider the model presented in \cite{ten}. This is
a GUT model with E$_6\times$U(1)$_{\rm X}$ gauge symmetry under which
the chiral superfields transform as 
\be
\Phi : \, \overline{351}_0 \; , \;\; \Psi_+ : \, 27_{+1} \; , \;\; 
\Psi_- : \; 27_{-1} ~.
\ee
The renormalizable superpotential, 
\be
W = \kappa\Phi\Psi_+\Psi_- ~,
\label{sup}
\ee
has a U(1)$_{\rm PQ}$ under which $\Phi$ has charge $-2$ and
$\Psi_+$, $\Psi_-$ have charge +1.
This is broken by dimension-6 and higher operators in the
superpotential:
\be
W_{\rm nr} = \frac{1}{M_{\rm P}^3} 
\left(\frac{\kappa_1}{6}\Phi^6 + 
\frac{\kappa_2}{3}
(\Psi_+ \Psi_-)^3 + \frac{\kappa_3}{4}\Phi^4\Psi_+\Psi_-\right) 
+ ...  
\ee
where the coefficients $\kappa_j$ are expected to be 
of order one.
As observed in ref.~\cite{dudas}, their interference with the 
renormalizable superpotential gives dimension-7 operators
in the Lagrangian:
\be
\frac{1}{M_{\rm P}^3} \Phi^5 \Psi_+^{\dagger} \Psi_-^{\dagger} \; , \;\; 
\frac{1}{M_{\rm P}^3} \Psi_{\pm}^5 \Psi_{\mp}^{\dagger} \Phi^{\dagger}
 \; , \;\; \frac{1}{M_{\rm P}^3} \Phi^4 \Phi^{\dagger} 
\left| \Psi_{\pm} \right|^2
 ~, 
\label{dim7}
\ee
where we use the same notation for the 
scalar components as for the corresponding chiral superfields.
The only fields which acquire vevs are the
scalar components of $\Phi$ (the Higgs). Therefore, according to the
arguments of section 2.1, the  operators (\ref{dim7})
do not affect the solution to the strong CP
problem because they involve the scalar components of $\Psi_+$ and
$\Psi_-$, which have no vevs. 
The lowest dimensional operator in the
supersymmetric Lagrangian formed only of the $\Phi$ scalars 
is $\Phi^{11}\Phi^{\dagger 5}$, and is given by the interference
of the $\Phi^6$ and $\Phi^{12}$ terms in the superpotential.

However, the situation changes when soft susy breaking terms
are introduced.
Consider the 
\be
\kappa^{\prime} m_s \Phi\Psi_+\Psi_-
\label{tril}
\ee
trilinear scalar term, where $\kappa^{\prime}$ is a dimensionless 
coupling constant, and $m_s$ is the mass scale of 
susy breaking in the supersymmetric standard model.
The exchange of a $\Psi_+$ and a $\Psi_-$ scalar between this operator
and the first operator in (\ref{dim7}) leads at one loop
to a six-scalar effective term in the Lagrangian:
\be
- \frac{2\kappa^*\kappa_1\kappa^{\prime}}{(4\pi)^2}
\frac{m_s}{M_{\rm P}^3} \log \left(\frac{M_{\rm P}}{m_s}\right) \Phi^6 ~.
\label{danger}
\ee
The constraint from $\bar{\theta}$ given by eq.~(\ref{first})
yields
\be
|\kappa\kappa_1\kappa^{\prime}| < {\cal O}(10^{-18}) ~,
\ee
where we have used $m_s \sim {\cal O}(250$ GeV). 
Note that there are also one-loop $\Phi^{6}\left|\Phi\right|^{2 n}$
terms which can be summed up. 
In addition, once soft susy breaking masses are introduced,
the unwanted five-scalar term
$ \Phi^4 \Phi^{\dagger}$ is induced at one-loop by contracting
the $\Psi$ legs of the 
third operator in (\ref{dim7}). This term is independent
of the trilinear soft term (\ref{tril}). 
Thus, the coupling constants that appear
in the renormalizable superpotential, 
in the soft terms, or in the non-renormalizable terms from the
superpotential have to be very small, contrary to the goal 
of this model.

In ref.~\cite{dudas} it is suggested that an additional chiral
superfield, $\Upsilon$, which transforms non-trivially under both
E$_6$ and U(1)$_{\rm X}$, may allow different $X$ charges for
the $\Phi$, $\Psi_1$ and $\Psi_2$ superfields while satisfying
the gauge anomaly cancellation and 
avoiding the dangerous PQ breaking operators.
We point out that $\Upsilon$ should transform in a real representation
of E$_6$ to preserve the (E$_6$)$^3$ anomaly cancellation.
The lowest real representation is the adjoint 78 and has index 2
in the normalization where the  fundamental 27 and the antisymmetric
351 have indices 1/2 and 25/2, respectively.
The gauge invariance of the renormalizable superpotential
(\ref{sup}) requires $X_{\Psi_1} + X_{\Psi_2} = - X_{\Phi}$,
which, together with the (E$_6$)$^2 \times $U(1)$_{\rm X}$
anomaly cancellation gives $X_{\Upsilon} = - 6 X_{\Phi}$.
Using these equations, we can write the (U(1)$_{\rm X})^3$
anomaly cancellation condition as a relation between the
U(1)$_{\rm X}$ charges of $\Psi_1$ and $\Psi_2$:
\be
(X_{\Psi_1} + X_{\Psi_2})
\left(X_{\Psi_1}^2 + \frac{407}{204} X_{\Psi_1}
X_{\Psi_2}  + X_{\Psi_2}^2 \right) = 0 ~.
\ee
The only real solution of this equation is $X_{\Psi_1} = - X_{\Psi_2}$
which does not prevent the dangerous operator (\ref{danger}).
The next real representation, 650, looks already too large
to allow a viable phenomenology.
Another proposal suggested in \cite{dudas} assumes fermion
condensation which now it is known not to occur in supersymmetric theories
\cite{susynp}.

%%%%%%%%%%%%%%%%%%%%%%%%%%%%%%%%%%%%%%%%%%%%%%%%%%%%%%
\subsection{The problem of PQ symmetry breaking scale}

If susy is relevant at the electroweak scale, and
susy breaking is transmitted to the fields of the standard model
by non-renormalizable interactions suppressed by powers of $M_{\rm
  P}$,
such as supergravity, then susy should be broken dynamically
at a scale $M_S \sim 10^{11}$ GeV. 
This will give scalar masses of order $M_W$. 
A gauge singlet in the dynamical susy breaking 
(DSB) sector with a vev for the $F$-term of order $M_S^2$ 
would produce gaugino masses of order $M_W$ \cite{ads}.
However, any gauge singlet is likely to have a mass of order
$M_{\rm P}$, so that its vev would need to be highly fine-tuned.
Nonetheless, gluino, neutralino
and chargino masses of order $M_W$ can be produced without need
for gauge singlets if there are new non-Abelian
gauge interactions which
become strong at $\sim$ 1 TeV \cite{gluino}.

The success of this scheme makes physics at the $M_S$ scale an
important candidate for spontaneously breaking a PQ-symmetry.
More important, the existence of $M_S$ in the rather narrow window
allowed for $f_{\rm PQ}$ is worth further exploration.
Nevertheless, models which break both susy and the PQ symmetry
face serious challenges, which were not addressed in the past
\cite{link}.
One obstacle is that the inclusion of colored fields in a model
of dynamical susy breaking 
typically results in a strongly coupled
QCD right above $M_S$ \cite{ads}. This problem can be solved 
by constructing a PQ sector that communicates with the DSB sector
only through a weak gauge interaction in a manner analogous
to the gauge mediated susy breaking models \cite{dns}.
A more serious problem is the following:
if colored superfields could be included in the
DSB sector, they would have masses of order $10^{11}$ GeV
and a non-supersymmetric spectrum. This will lead to large masses
for the squarks, which in turn will destabilize the electroweak scale.

The troublesome colored fields from the DSB sector 
can be avoided if the fields of both the DSB sector and 
the visible sector transform under the same global U(1)$_{\rm PQ}$, 
which is 
spontaneously broken in the DSB sector and explicitly broken in the 
visible sector by the color anomaly.
As pointed out in ref.~\cite{relax}, this may be possible because
the axion can be identified with one of the complex phases from
the soft susy breaking terms of the supersymmetric standard model.
However,
it appears very difficult to protect the axion against 
 Planck scale effects. For example, the $\mu$ and $B$ terms break this
U(1)$_{\rm PQ}$ which means that they should be generated by the vevs
of PQ-breaking products of fields from the DSB sector.
These products of fields are gauge invariant and therefore 
 Planck scale induced PQ-breaking operators may be induced.
The naturalness of the axion solution is preserved provided
these operators are suppressed by many powers of $M_{\rm P}$,
which in turn requires the vevs from the DSB sector to be much above
$M_S$ in order to generate large enough $\mu$ and $B$ terms.
Thus, this situation seems in contradiction with the cosmological
bounds on the PQ scale.
It should be mentioned though that a larger 
\pq\ might be allowed in certain unconventional cosmological 
scenarios suggested by string theory \cite{string}.
Note, however, that for larger \pq\ the constraints on PQ-breaking 
operators are significantly stronger [see eq.~(\ref{first})].

Another possibility, discussed in refs.~\cite{cla}, is to 
relate \pq\ to the susy breaking scale from the visible sector.
The idea is to induce negative squared masses at one-loop for some 
scalars, and to balance these soft susy breaking mass terms
against some terms 
in the scalar potential coming from the superpotential which are
suppressed by powers of $M_{\rm P}$.
By choosing the  dimensionality of these terms, one can ensure  
that the minimum of the potential is in the  range allowed for \pq.
This mechanism is also very sensitive to Planck scale effects
because it assumes the absence of certain gauge invariant
PQ breaking operators of dimension one, two and three from the 
superpotential.

Given these difficulties in 
relating \pq\ to the susy breaking scale while avoiding
the harmful low-dimensional operators,
one may consider producing the PQ scale naturally by introducing
some gauge interactions which become strong at about $10^{11}$ GeV
and break the PQ symmetry without breaking susy.
Because this scenario is less constrained, it may be easier to avoid
the PQ-breaking operators.

%%%%%%%%%%%%%%%%%%%%%%%%%%%%%%%%%%%%%%%%%%%%%%%%%%%%%%%%%%%%%%%

%%%%%%%%%%%%%%%%%%%%%%%%%%%%%%%%%%%%%%%%%%%%%%%%%%%%%%%%%%%%%%%
\section{Conclusions}
\label{sec:conc}

We have argued that an axion model has to satisfy two naturalness
conditions in order to solve the strong CP problem: 

\noindent
a) the absence of low-dimensional
Planck-scale induced PQ-breaking operators
formed of fields which acquire vevs;

\noindent
b) the absence of fundamental mass parameters much smaller than 
the Planck scale.

If these conditions are not satisfied, the models may 
not be ruled out 
given that the Planck scale physics is unknown, but
the motivation for the axion models (i.e. avoiding fine-tuning) is
lost.

Non-supersymmetric composite axion models satisfy condition b)
easily. The only phenomenological problem is that they predict heavy
stable particles which are ruled out by the thermal evolution of the
early universe. However, this problem disappears if there is inflation
with reheating temperature below the PQ scale.
Condition a) is more troublesome. It is satisfied  by
only one composite axion model \cite{model}, and our estimate shows
that it leads to a Landau pole for QCD. One may hope though that
the uncertainty in the value of $M_{\rm P}$, i.e. the possibility
of quantum gravitational effects somewhat below $10^{19}$ GeV,
combined with unknown non-perturbative effects of axicolor
on the running of the 
strong coupling constants, might push the Landau pole just above $M_{\rm
  P}$ where is irrelevant for field theory.
But because this does not seem to be a probable scenario,
it would be useful to study in detail the 
possibility of unifying color with axicolor.

By contrast, the existing supersymmetric models do not satisfy condition
a).
The models which attempt to eliminate the PQ-breaking operators
rely on the holomorphy of the superpotential. We have shown that
once susy breaking is taken into account, the PQ-breaking operators
are reintroduced with sufficiently large coefficients (in the absence
of fine-tuning) to spoil the solution to the strong CP problem.
Also, the models that satisfy condition b) by relating the PQ scale
to the susy breaking scale are particularly sensitive 
to gauge invariant PQ-breaking operators.
These results suggest the need for further model building efforts.

%%%%%%%%%%%%%%%%%%%%%%%%%%%%%%%%%%%%%%%%%%%%%%%%%%%%%%%%%%%%%%%
\section*{Acknowledgements}

I am grateful to Sekhar Chivukula for many helpful discussions
about strongly coupled theories and axions.
I would like to thank Lisa Randall for very useful observations
on the manuscript. I also thank Indranil Dasgupta, 
Ken Lane, Martin Schmalz and John Terning for useful discussions,
and Emil Dudas, Tony Gherghetta and Scott Thomas for 
valuable correspondence.

{\em This work was supported in part by the National Science
  Foundation under grant PHY-9057173, and by the Department of Energy
  under grant DE-FG02-91ER40676.}

%%%%%%%%%%%%%%%%%%%%%%%%%%%%%%%%%%%%%%%%%%%%%%%%%%%%%%%%%%%%%%%

\vfil
\end{document}